\documentclass[epj]{svjour}  
\usepackage{amssymb}  
\usepackage{graphics}  
\usepackage{pdfpages}  
  
\begin{document}  
\title{Excitation of the electric pygmy dipole resonance by 
inelastic electron   scattering}  
  
\author{V.Yu.~Ponomarev\inst{1},  
D.H.~Jakubassa-Amundsen\inst{2}, 
A. Richter\inst{1}, and
J. Wambach\inst{1,3}  
}                     
  
\institute{  
Institut f\"{u}r Kernphysik, Technische Universit\"{a}t  
Darmstadt, D-64289 Darmstadt, Germany  
\and   
Mathematisches Institut, Universit\"{a}t   M\"{u}nchen,  
Theresienstra\ss e 39, D-80333 M\"{u}nchen, Germany   
\and
European Centre for Theoretical Studies in Nuclear Physics and 
related Areas (ECT*) and Fondazione Bruno Kessler, Villa
Tambosi, 38123 Villazzano (TN), Italy
}  
\date{Received: date / Revised version: date}  
  
%
%
%
\abstract{  
To complete earlier studies of the properties of the electric pygmy 
dipole resonance (PDR) obtained in various nuclear reactions, the 
excitation of the 1$^-$ states in $^{140}$Ce by $(e,e')$ scattering for 
momentum transfers $q=0.1-1.2$~fm$^{-1}$ is calculated within the 
plane-wave and distorted-wave Born approximations. The excited states 
of the nucleus are described within the Quasiparticle Random Phase 
Approximation (QRPA), but also within the Quasiparticle-Phonon Model 
(QPM) by accounting for the coupling to complex configurations.
It is demonstrated that the excitation mechanism of the PDR states in
$(e,e')$ reactions is predominantly of transversal nature for scattering 
angles $\theta_e \approx 90^o-180^o$. Being thus mediated by the 
convection and spin nuclear currents, the $(e,e')$ like the 
$(\gamma,\gamma')$ reaction, may provide additional information to 
the one obtained from Coulomb- and hadronic excitations of the PDR in
$(p,p')$, $(\alpha,\alpha')$, and heavy-ion scattering reactions.  
The calculations predict that  the $(e,e')$ cross sections for the
strongest individual PDR states are in general about three orders of 
magnitude smaller as compared to the one of the lowest $2^+_1$ state 
for the studied kinematics, but that they may become dominant at 
extreme backward angles.
\PACS{  
      {25.30.Dh}{Inelastic electron scattering to specific states}  
     } 
} 
\authorrunning{V.Yu. Ponomarev {\it et al.}}  
\titlerunning{The PDR excitation in $(e,e')$ scattering}  
\maketitle  
%

{\it Dedicated to Pier Francesco Bortignon}

\section{Introduction}  
  
A group of low-lying $1^-$ states in neutron-rich heavy nuclei below the 
particle 
emission threshold is often referred to as the Pygmy Dipole Resonance (PDR). 
The excitation probability of the PDR by photons is about two 
orders of magnitude smaller as compared to the Giant Dipole Resonance (GDR).  
Nonetheless the high selectivity of the electromagnetic interaction to the 
excitation of dipole states already allowed to observe the PDR in experiments with 
tagged photons as a bump of unresolved states with a width of about 2-3~MeV
\cite{Lasz1,Lasz2}.
Later, nuclear resonance fluorescence (NRF) experiments with high
resolution in Darmstadt \cite{He97} and Gent \cite{Go98} 
and follow-up studies during the last 20 years 
also at the ELBE accelerator
of the Helmholtz-Zentrum Dresden-Rossendorf
\cite{ELBE}
and at the High Intensity $\gamma$-ray Source 
(HI$\gamma$S) operated by 
Triangle University Nuclear Laboratory (TUNL) \cite{TUNL},
identified the PDR fine structure, i.e. hundreds of 1$^-$ states 
were observed in spherical nuclei at the PDR excitation energy.

Recently, other probes were used to investigate the PDR properties: $1^-$ states which
form the PDR, were studied in $(\alpha,\alpha'\gamma)$ \cite{aa}, $(p,p')$
\cite{pp}, ($^{17}$O,$^{17}$O$'\gamma$) \cite{17O}, and $(p,p'\gamma)$
\cite{Ce-last} reactions, in which the detection of the  $\gamma$ decay 
photon
in coincidence with the scattered particle was used to
select the corresponding excitation of a $1^-$ state.
For example, the spectrum of the $1^-$ states in $^{208}$Pb obtained in  the
$(p,p')$ reaction at very small scattering angles ($\theta_{\rm lab} < 1^o$, 
where the excitation process is purely determined by the
Coulomb interaction between projectile and target
\cite{pp}) resembles closely the NRF spectrum
\cite{Rye}.
At the same time, Coulomb- and strong ($NN$) interactions
between projectile 
and nucleus play an important role in the
excitation of the PDR states in the other reactions mentioned above.
As a result of the different sensitivity of various reactions, 
some $1^-$ states are observed in NRF spectra but not in
reactions with hadronic probes and vice versa. Also, the relative excitation
strengths of different individual states deviate appreciably.
Contrary to the GDR states where the $E1$-strength is concentrated 
merely in a single collective level called 1p-1h doorway state
and a spreading over many states of 2p-2h, 3p-3h, $\ldots$ character (see,
e.g., \cite{BBB}), the PDR is characterized by probably a few doorway
states. We will return to this point in some detail below. 

For a detailed account of the present status of studies of the PDR
properties we refer to a recent review article \cite{rew}.

In the present work we consider the possibility of using electrons as a
projectile to supply further information on the properties of the $1^-$ states
belonging to the PDR.  
As it will become clear from results presented below,
the prerequisite for an experimental verification of them are the
availability of (i) low energy electron beams and (ii) high-resolution and
large acceptance magnetic spectrometers.
Both conditions are, e.g., fulfilled at the S-DALINAC ({\bf S}uperconducting
{\bf Da}rmstadt Electron {\bf Lin}ear {\bf Ac}celerator) 
and its spectrometers LINTOTT and Q-CLAM \cite{i}.
Some selective excitations of isoscalar and isovector electric dipole
transitions below the electric giant resonance region were, e.g.,
investigated in $^{12}$C, $^{16}$O, $^{40}$Ca and $^{208}$Pb
\cite{ii,iii,iv,v}.
Furthermore, some benchmark high-resolution $(e,e'x)$ 
experiments with $x = p, n, \alpha$
and the decay of the Giant Dipole Resonance (GDR) in the doubly
magic nuclei $^{40}$Ca and $^{48}$Ca were also performed at the S-DALINAC
\cite{vi,vii,viii,viv,vv} but in general, 
the information on the observation  of detailed strength distributions of 
1$^-$ states in the $(e,e')$ reactions 
is very sparse.

Concerning the physical origin of low energy electric dipole strength and
its particular distribution we note in passing that there exists at present
still no clear picture about the relevant excitation mechanism.  
Recent self-consistent Random-Phase Approximation (RPA) calculations with
various finite-range forces in $^{16}$O and $^{40}$Ca \cite{Papa1} 
and also $^{48}$Ca \cite{Papa2} have shown that, e.g., nuclear surface
vibrations might mix with skin modes and thus influence the pygmy dipole
strength.
It is stated clearly there that an electroexcitation experiment of the
$(e,e')$ type could eventually help to``improve the different models
aspiring to describe reliably the low-energy dipole strength of nuclei"
\cite{Papa2}. 
This point has also been independently emphasised in \cite{Rei}. 
To provide some estimates for the feasibility of $(e,e')$
experiments is the main purpose of this article.

The cross section for the excitation of natural parity states in 
$(e,e')$ reactions has a longitudinal and transversal component.
It is expected that the longitudinal  or Coulomb term  gives rise to a
distribution of electric dipole strength over energy quite similar to the
one seen in NRF experiments, at least at small momentum transfer.
However, the transversal part is mediated by nuclear currents, and thus provides
an alternative mechanism to excite the same set of PDR states in addition to
the Coulomb and $NN$ excitations.

It is thus important

-- to investigate at which kinematics the transversal mechanism dominates
over the longitudinal one in the excitation of the PDR states, and to compare it
to the behaviour of the excitation of the collective GDR and

-- to provide realistic estimates of the $(e,e')$ cross
section for the excitation of the PDR levels. 

Electrons with incident kinetic energies from 30 to 120~MeV will be considered.   
Such energies can be provided by the S-DALINAC in Darmstadt, where the detector system allows for 
measurements in a wide range of scattering angles, including 
backward scattered electrons close to 180$^o$ which can be
detected with high angular resolution \cite{i,Lut}.

The calculations have been performed for $^{140}$Ce, a semi-magic 
nucleus in which the PDR
has  already been studied in $(\gamma,\gamma')$, $(p,p')$, and
$(\alpha,\alpha')$ reactions \cite{Ce-last}.
  
\section{Plane-wave Born approximation}  
  
The theory of inelastic scattering of electrons on nuclei is well 
developed and may be found in textbooks (see, e.g., \cite{Ube}).   
The plane-wave Born approximation (PWBA) is   
usually sufficient for simple estimates. In the PWBA, the   
differential $(e,e')$ cross section for excitation of a natural parity state  
of multipolarity $\lambda$ can be written as \cite{HB83}
\begin{equation}  
\left(\frac{d\sigma}{d\Omega}\right)_{\lambda} \propto \left\{  
V_L(\theta_e)\left|F^C_{\lambda}(q)\right|^2  
+ V_T(\theta_e)\left|F^E_{\lambda}(q)\right|^2 \right\}  
\label{eq1}  
 \end{equation}  
where $V_L(\theta_e)$ and $V_T(\theta_e)$ are the longitudinal and  
transversal kinematic factors, respectively, and $\lambda$ denotes the
multipolarity of the transition.
Nuclear structure information  
on the excited state enters via the charge transition density $\rho_{\lambda}(r)$   
into the Coulomb form factor
\begin{equation}  
F^C_{\lambda}(q) \propto \int_0^{\infty}\rho_{\lambda}(r) j_{\lambda}(qr)r^2  
dr  
\label{eq2}  
\end{equation}  
and via the transition current densities $J_{\lambda,\lambda\pm1}(r)$ into the  
electric form factor
\begin{eqnarray}  
F^E_{\lambda}(q) &\propto& \int_0^{\infty}  
\left\{\sqrt{\lambda+1}~J_{\lambda,\lambda-1}(r)~ j_{\lambda-1}(qr) \right .  
\nonumber  
\\  
&+&  
\left .  
\sqrt{\lambda}~J_{\lambda,\lambda+1}(r) ~j_{\lambda+1}(qr)  
\right\}  
r^2 dr  
\label{eq3}  
\end{eqnarray}  
where $q$ denotes the three-momentum transfer and $j_{\lambda}(qr)$ is the spherical   
Bessel function. Any interference between Coulomb and electric form factors is  
neglected in the PWBA.   
  
\begin{figure}  
\begin{center}  
\includegraphics[width=5.5cm]{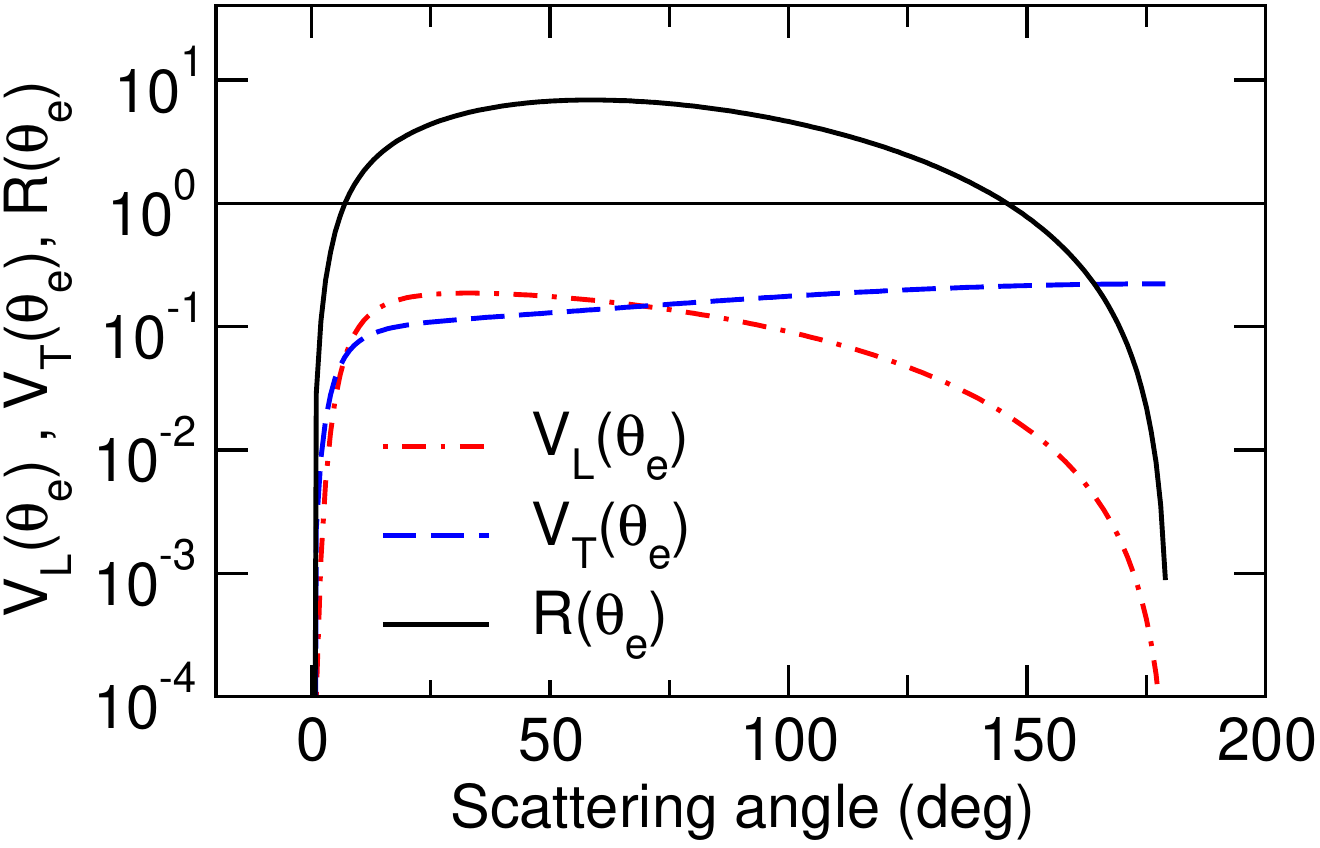}  
\end{center}  
\caption{\label{Fig1}  
Kinematic factors $V_L(\theta_e)$ and $V_T(\theta_e)$ in Eq.~(\ref{eq1})  
and the quantity R($\theta_e$) in Eq.~(\ref{eq5}) as function of the scattering angle 
$\theta_e$ for electrons with incident energy $E_e = 70$~MeV.  
See text for details.}  
\end{figure}

At small $q$-values, Siegert's theorem~\cite{Sie} may be applied,   
resulting in
\begin{equation}  
F^E_{\lambda}(q) \approx \frac{E_x}{q}\sqrt{\frac{\lambda+1}{\lambda}}  
~F^C_{\lambda}(q)   
\label{eq4}  
\end{equation}  
where we have used relativistic units ($\hbar=c=1$).  
When combining Eqs. (\ref{eq1}) and (\ref{eq4}), the quantity  
\begin{equation}  
{\rm R}(\theta_e) = \frac{V_L(\theta_e)}{\frac{E_x}{q}\sqrt{\frac{\lambda+1}  
{\lambda}}  
~V_T(\theta_e)}  
\label{eq5}  
\end{equation}  
indicates whether the longitudinal or the transversal contribution 
dominates in the nuclear excitation process. This quantity is shown in  
Fig.~\ref{Fig1} by a solid line together with the kinematic factors  
$V_L(\theta_e)$   
and $V_T(\theta_e)$. The calculation was performed for a hypothetical
$1^-$ state with an excitation
energy of $E_x = 8$~MeV excited by $E_e = 70$~MeV electrons.   
Notice that the excitation via the Coulomb form factor   
(${\rm R}(\theta_e) > 1$)  
dominates in a wide range of  scattering angles  
$\theta_e$,  
except for very forward ($\theta_e = 0^o-8^o$) and backward ($147^o-180^o$)  
angles.  
  
When the momentum transfer $q$ is small, it is also possible to perform a   
Taylor expansion of the Bessel function $j_{\lambda}(qr)$ in Eq.~(\ref{eq2}).  
Keeping only the first term, the square of the Coulomb form factor is  
closely related to the reduced transition probability $B(E \lambda)$ of   
the excited state,  
$|F^C_{\lambda}(q)|^2 \propto q^{2\lambda}~B(E\lambda)$.  
  
These simple estimates lead to the expectation that the distribution of $E1$
strength of states in the region of the   
PDR in $(e,e')$ experiments at small $q$-values is rather similar to the   
one in $(\gamma,\gamma')$ measurements. Indeed, at fixed kinematics, the 
$(\gamma,\gamma')$ excitation cross section is strictly proportional to  
$B(E\lambda)$ (see, e.g., \cite{Go94,wf}).    
Some deviations are possible only at very large scattering angles of
electrons.  
  
The nuclear structure information on the states  
which form the PDR is contained in   
transition charge and current densities, which enter  
into Eqs.~(\ref{eq2}) and (\ref{eq3}). They were calculated 
for $^{140}$Ce within the   
quasiparticle-phonon model~\cite{Solo,LoIu}.  
The model employs a nuclear Hamiltonian which includes the mean field for  
protons and neutrons (a phenomenological Woods-Saxon potential is usually used),  
monopole pairing, and residual interactions in a separable multipole form.    
Excitations of even-even nuclei are treated as quasi-bosons (phonons),  
the excitation energies and internal fermion structure of which are obtained   
by solving equations of motion of the quasiparticle random phase approximation (QRPA).   
This yields the eigenenergies and wavefunctions the one-phonon states.  
  
\begin{figure}  
\begin{center}  
\includegraphics[width=8.8cm]{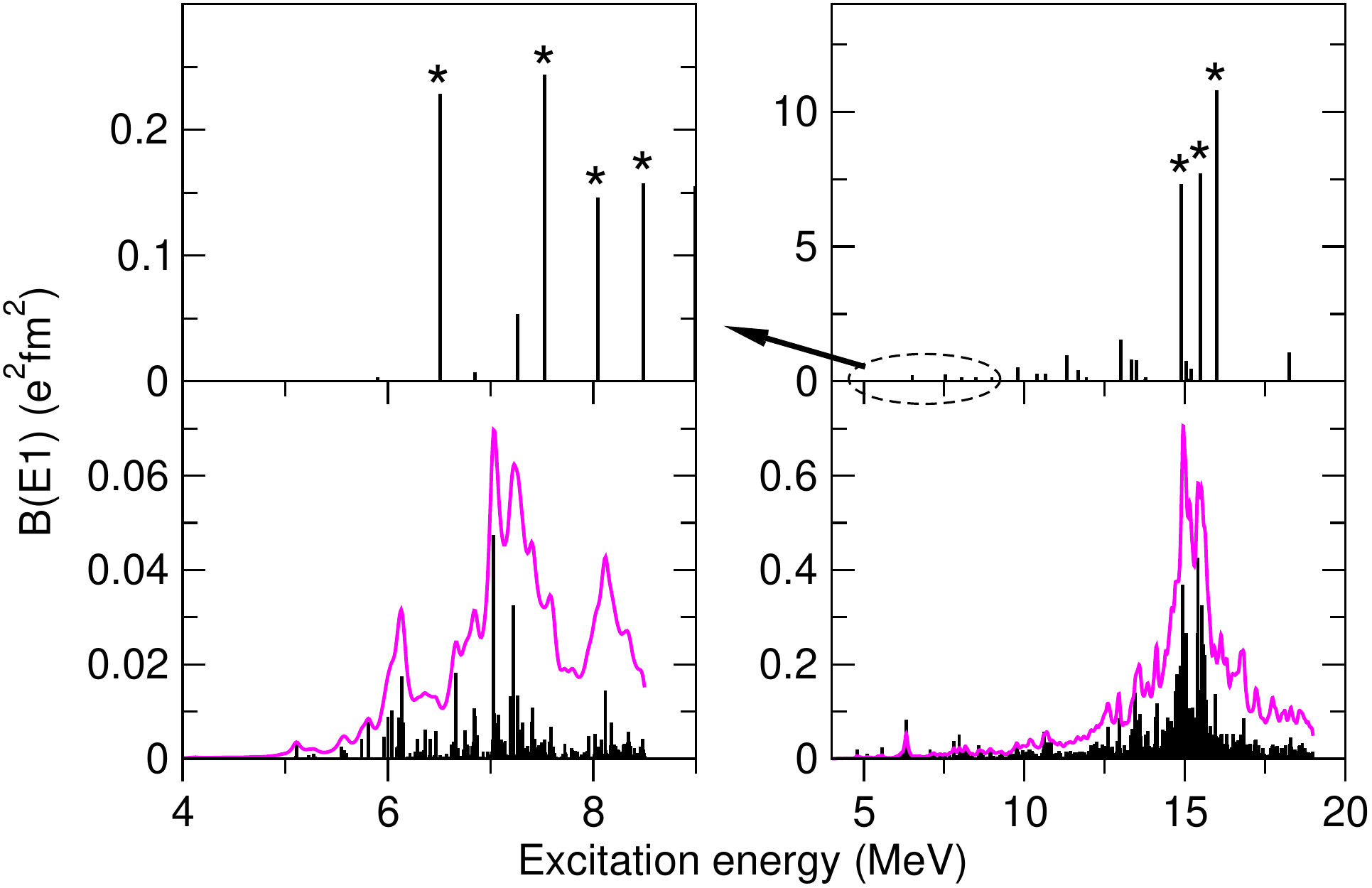}  
\end{center}  
\caption{\label{Fig2}  
The QPM prediction for the $B(E1)\uparrow$ strength distribution in $^{140}$Ce  
in the one-phonon approximation (top row) and  accounting for the coupling  
to complex configurations: two-phonon (bottom right) and two- and three-phonon 
ones (bottom left). See text for details.
The smooth curves 
in the bottom part (in relative units to guide 
the eye) are the result of an averaging   
over all states with a smearing parameter $\Gamma~=~0.1$~MeV (see  Eq.~(\ref{sf})).  
Selected one-phonon states for the discussion in the main text 
are marked by an asterix.  
}  
\end{figure}

The distribution of the $B(E1)$ strength over the one-phonon $1^-$ states   
in $^{140}$Ce in the PDR energy region is shown in Fig.~\ref{Fig2}~(top-left).   
The states with the largest $B(E1)$ values are marked with an asterix. 
They will be  
discussed in more detail below. Notice that their $B(E1)$ values are almost  
two orders of magnitude smaller as compared to the one-phonon states which   
form the GDR in Fig.~\ref{Fig2}~(top-right).    
  
\begin{figure}  
\begin{center}  
\includegraphics[width=8.5cm]{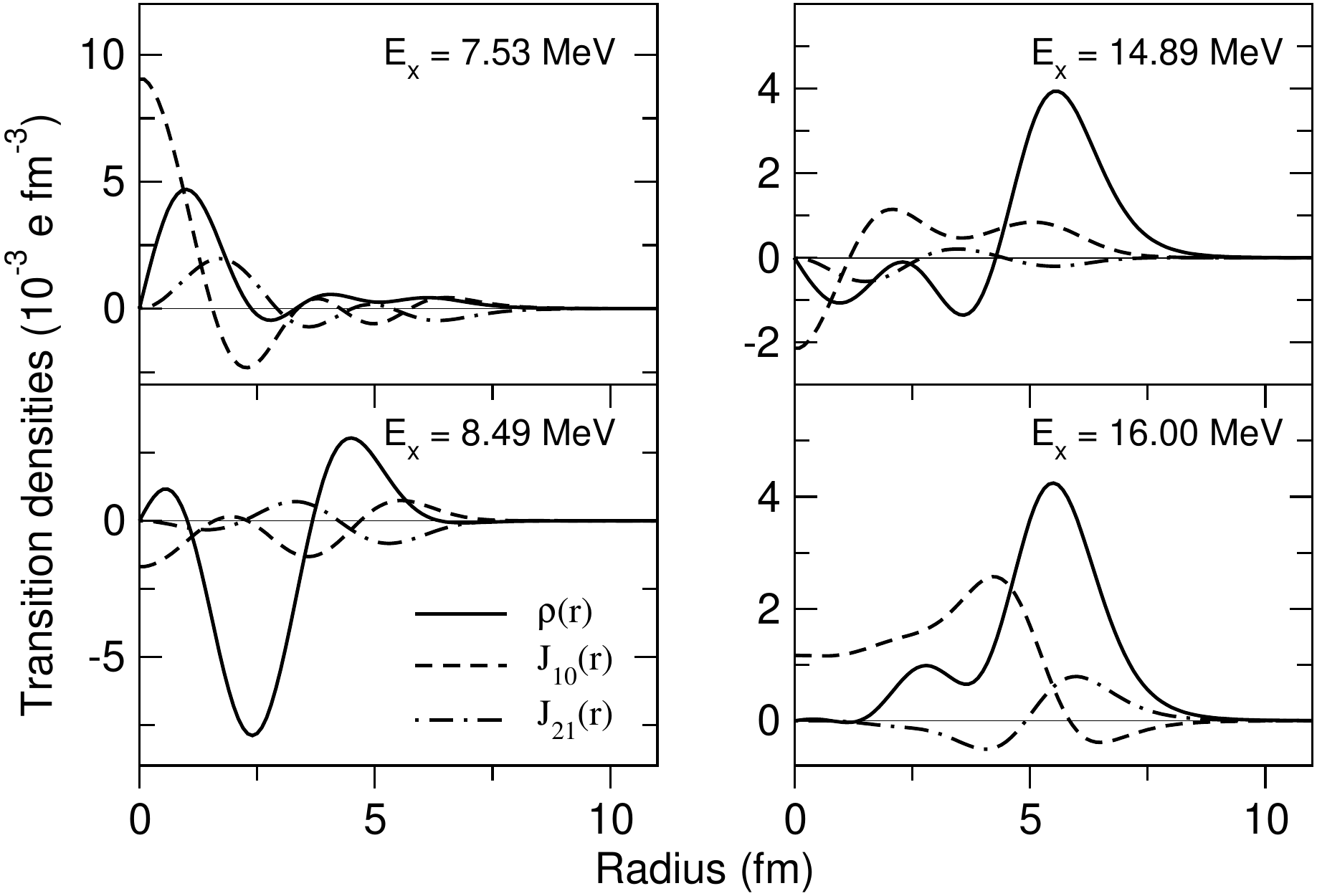}  
\end{center}  
\caption{\label{Fig3}  
Transition charge $\rho_1(r)$ (solid line) and   
current $J_{1,0}(r)$ (dashed line) and $J_{1,2}(r)$ (dash-dotted line)  
densities   
as a function of the radial coordinate $r$ for the excitation of  
some one-phonon $1^-$ states in $^{140}$Ce.}  
\end{figure}

Transition charge density, $\rho_1(r)$, and current densities, 
$J_{1,0}(r)$ and $J_{1,2}(r)$,
of some selected one-phonon $1^-$ states are presented in  
Fig.~\ref{Fig3} with their excitation energies indicated in the right-top   
corner of each panel. The ones belonging to the PDR (GDR) are shown in the   
left (right) column.  
The following effective charges for protons (Z) and neutrons (N) have been 
used: $e_{Z(N)} = {\rm N(-Z)/A}$  
for the $B(E1)$ values and charge densities, and the effective $g$-factors:  
$g_l^{Z(N)}= e_{Z(N)}$ for the convection  
current, and  $g_s^{\rm eff} = 0.8~g_s^{\rm free}$ for the magnetization  
current.   
  
The charge transition densities of the $1^-$ states, which form the GDR, have a strong  
surface peaking, typical for collective vibrations. Protons and neutrons  
oscillate out of phase and due to different signs of the effective charges,   
they add constructively. The interference of 1p1h components in the wave function of  
one-phonon $1^-$  states from the PDR energy region has a destructive  
nature~\cite{Pon}. Their charge transition densities are  
peaking in the interior of the nucleus where their main   
1p1h component is dominating. Accordingly, the position of minima and 
maxima varies from state to state.      
  
\begin{figure}  
\begin{center}  
\includegraphics[width=8.5cm]{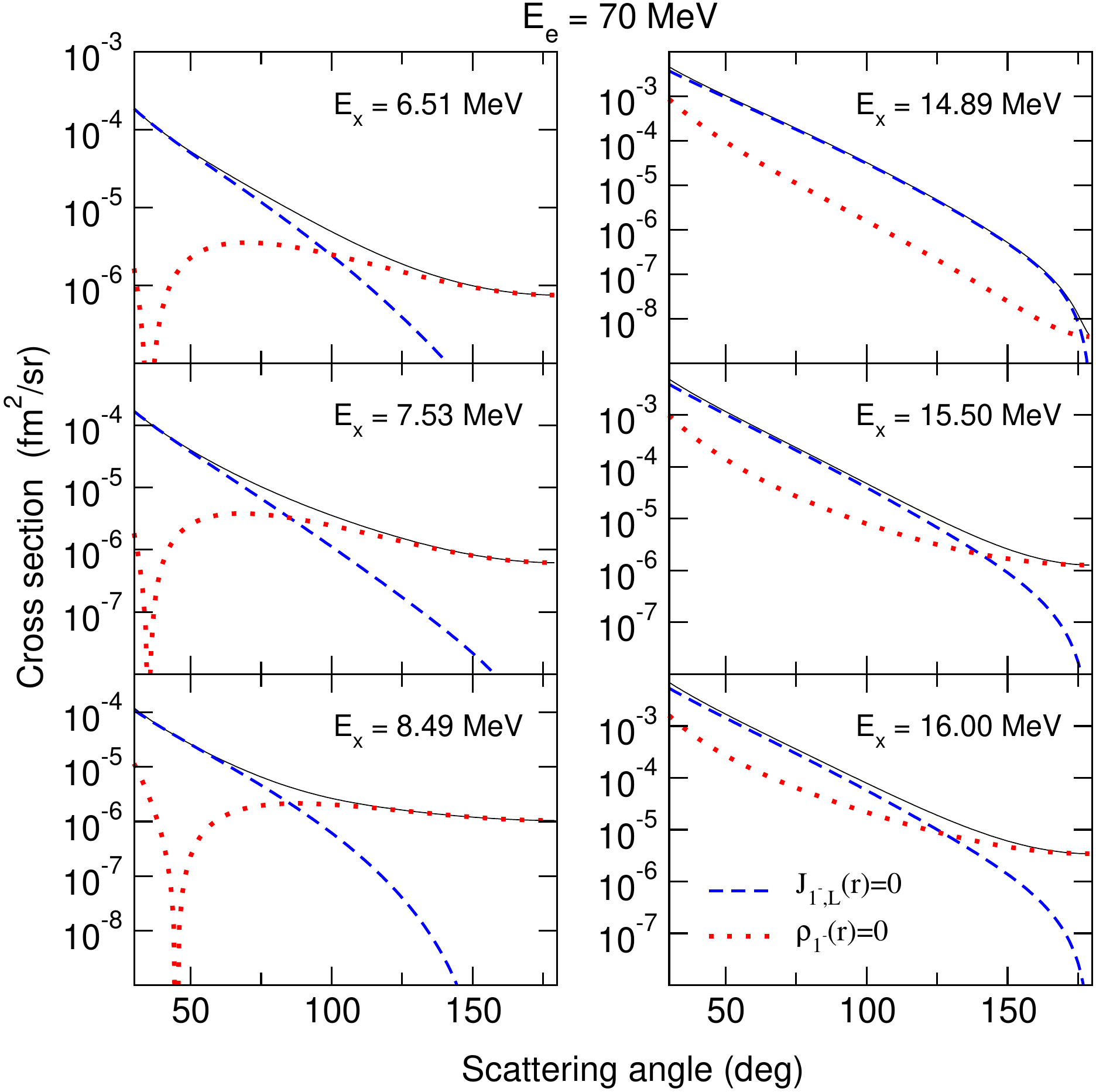}  
\end{center}  
\caption{\label{Fig4}  
Differential PWBA cross sections (d$\sigma$/d$\Omega$) for the excitation  
of one-phonon $1^-$ states from the PDR (left  
column) and the GDR (right column) energy region in $^{140}$Ce as a function of 
scattering angle $\theta_e$. The  incident energy is 70~MeV.  
The longitudinal and transversal components are displayed as dashed and dotted   
lines, respectively. The total cross sections are shown by solid lines.  
}  
\end{figure}

The PWBA differential $(e,e')$ cross sections for the excitation of three selected  
one-phonon $1^-$ states, which belong to the PDR (GDR) 
are presented in the left (right) part of Fig.~\ref{Fig4} as a function 
of the angle of the scattered electrons.  
They are calculated for an incident energy of 70~MeV. The contribution of the  
longitudinal and transversal components is shown separately by dashed and  
dotted lines, respectively. 
Our conclusion about  
the longitudinal and the transversal contributions based on Eq.~(\ref{eq5}) 
and Fig.\ref{Fig1} remains 
valid for the GDR states. But for the PDR states the transversal component 
determines the excitation  
in a wide range of angles from $90^o$ to $180^o$. For all three  
PDR states the differential cross section has a very similar 
shape with deep minima between $40^o$ and $50^o$ and  
an almost flat behaviour for scattering angles $\theta_e > 70^o$.     
  
\section{Distorted-wave Born approximation}  
  
Within the distorted-wave Born approximation (DWBA), one solves the Dirac   
equation for the incoming and outgoing electrons in the Coulomb field of the nucleus 
in terms of partial waves.  
Schematically, the differential cross section has the form:  
\begin{equation}  
\left(\frac{d\sigma}{d\Omega}\right)_{\lambda} \propto  
\sum_{m_s,\mu} \left|  
A(\lambda \mu m_s)  
\right|^2  
\label{eq6}  
\end{equation}  
where $m_s$ and $\mu$ are the projections of the spin $\vec{s}$   
of the incoming electron   
and of the angular momentum transfer $\vec{\lambda}$, respectively.  
The expression for the transition amplitudes $A(\lambda \mu m)$   
may, e.g., be found in \cite{Ube,Tuan}.   
An essential detail is that $A = A_C + A_E$, where both Coulomb ($A_C$)  
and electric ($A_E$) amplitudes are calculated by folding  the nuclear  
charge and current transition densities, respectively, with the partial waves  
of the incoming and outgoing electron and with the propagator of the virtual photon.  
This implies that the DWBA accounts for the interference between two mechanisms  
for the excitation of natural parity states, (longitudinal) Coulomb and 
(transversal) electric.  
  
One of the DWBA problems is a poor convergence of the radial integrals,  
particularly for dipole excitations at backmost scattering angles.  
We employ here the complex-plane rotation method developed 
in~\cite{VF70,we1}, to  overcome this problem.   
This allows us to cover all scattering angles from $0^o$ to $180^o$.  
The computation time is sped up by a multiple convergence acceleration in the 
sum over the final-state partial waves \cite{YRW54}. However, 
such an acceleration is not possible for angles $\theta_e \lesssim 10^\circ$.  
  
\begin{figure}  
\begin{center}  
\includegraphics[width=8.5cm]{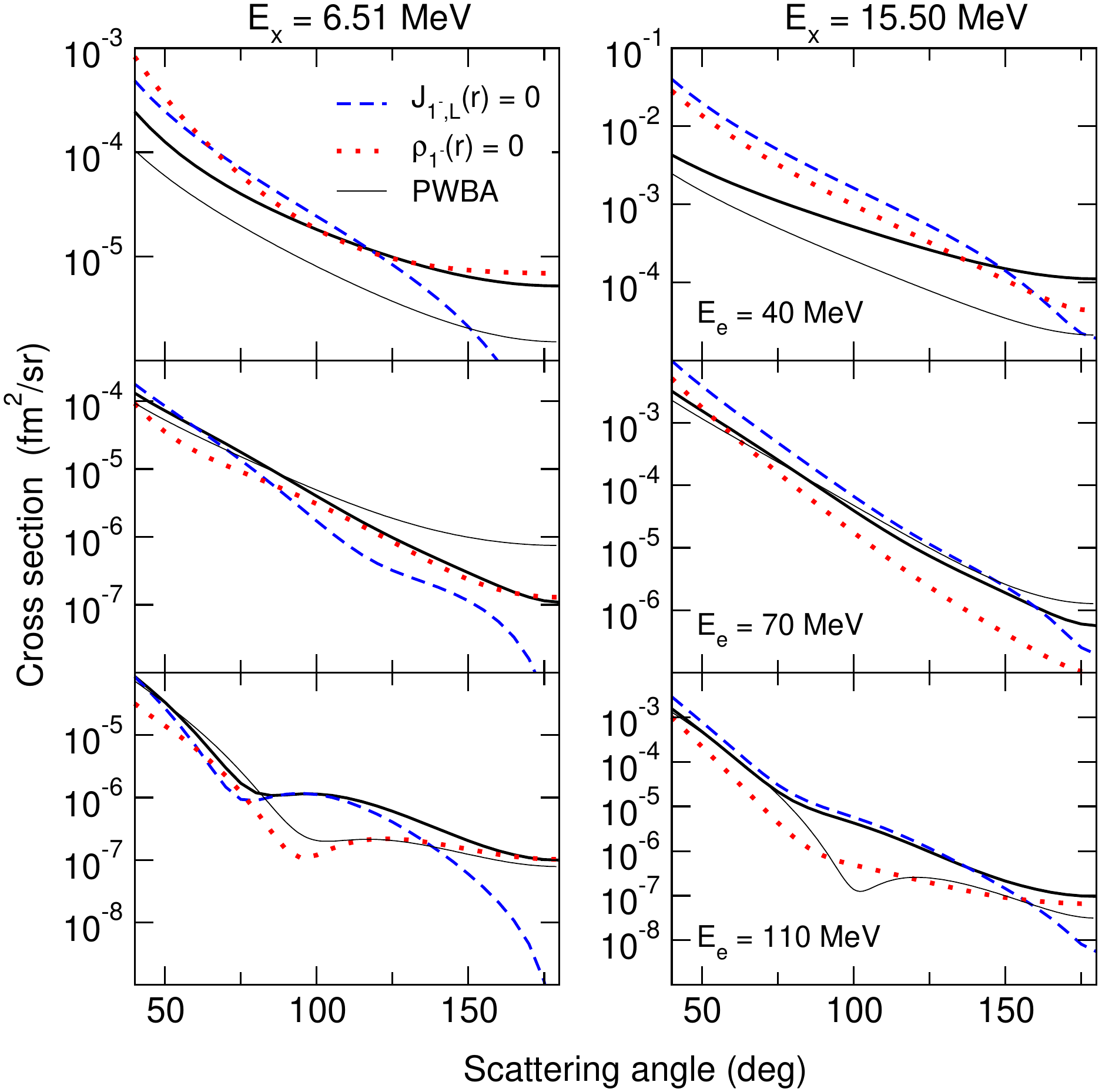}  
\end{center}  
\caption{\label{Fig5}  
Differential cross sections (d$\sigma$/d$\Omega$) for the excitation  
of the one-phonon $1^-$ states of energy 6.51~MeV (left  
column) and 15.50~MeV (right column) in $^{140}$Ce as a function of scattering   
angle $\theta_e$. The incident energy is 40~MeV (top row), 70~MeV (middle row),  
and 110~MeV (bottom row).  
The DWBA and PWBA results are represented by solid thick and thin lines,  
respectively.   
The results of the DWBA calculation in which only a ``longitudinal" or  
``transversal" excitation is accounted for are displayed as  dashed and dotted   
lines, respectively.  
}  
\end{figure}

\begin{figure}  
\begin{center}  
\includegraphics[width=8.5cm]{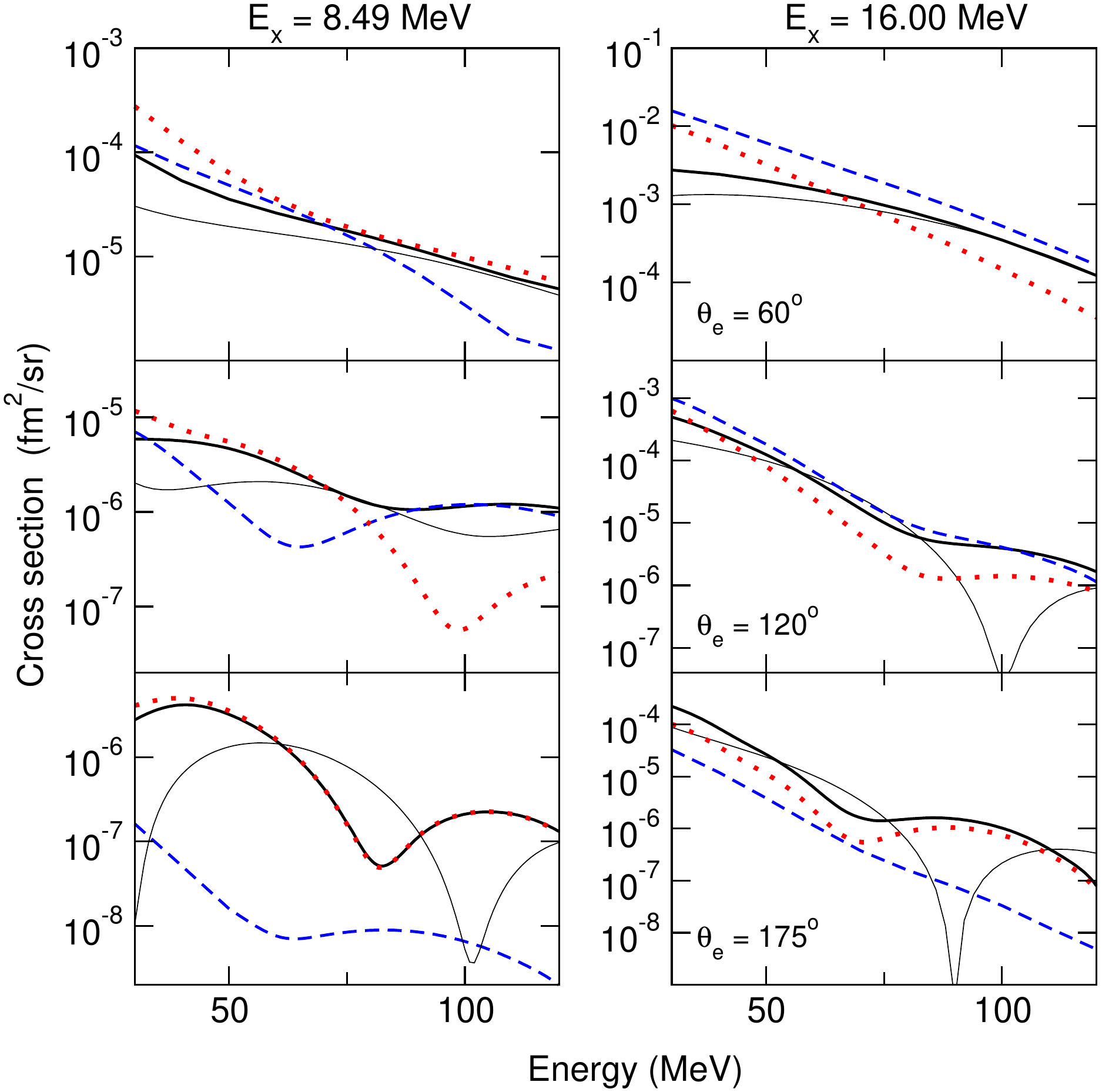}  
\end{center}  
\caption{\label{Fig6}  
Differential cross sections (d$\sigma$/d$\Omega$) for the excitation  
of the one-phonon $1^-$ states of energy 8.49~MeV (left  
column) and 16.00~MeV (right column) in $^{140}$Ce at a scattering  
angle $60^o$ (top row), $120^o$ (middle row), and $175^o$ (bottom row)  
as a function of kinetic energy $E_e$ of the  electrons.  
The definition of the lines is the same as in Fig.~\ref{Fig5}.  
}  
\end{figure}

\begin{figure}  
\begin{center}  
\includegraphics[width=8.5cm]{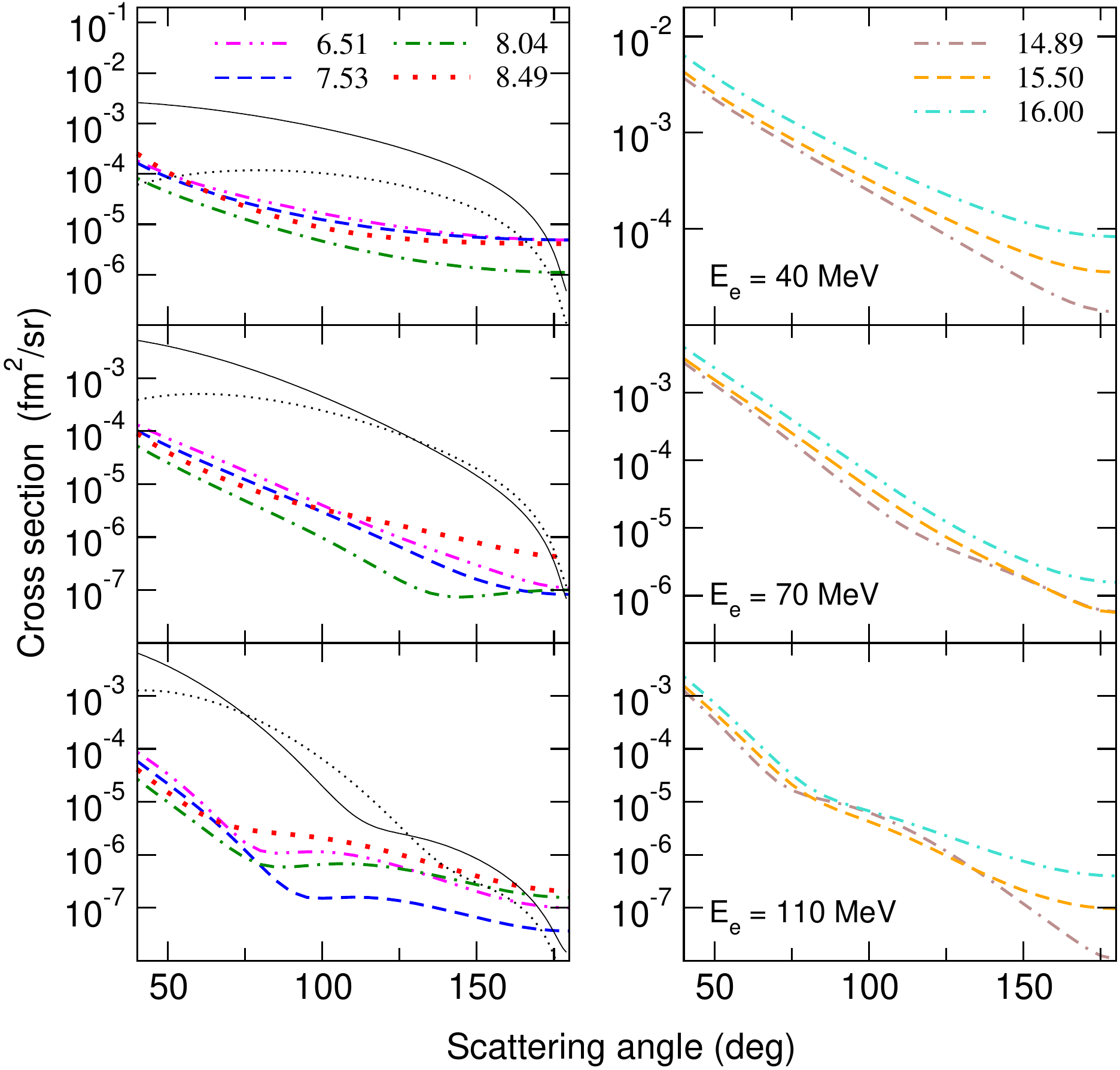}  
\end{center}  
\caption{\label{Fig7}  
Differential cross sections (d$\sigma$/d$\Omega$) for the excitation  
of the selected one-phonon $1^-$ states from the PDR (left  
column) and GDR (right column) region in $^{140}$Ce as a function of scattering   
angle $\theta_e$. The incident energy is 40~MeV (top row), 70~MeV (middle row),  
and 110~MeV (bottom row).  
Excitation energies of the states are given in the inserts of the top row.  
Cross sections for the excitation of the $2^+_1$ (solid line) and  
$3^-_1$ (thin dotted line) states are plotted for comparison.  
}  
\end{figure}

\begin{figure}  
\begin{center}  
\includegraphics[width=8.5cm]{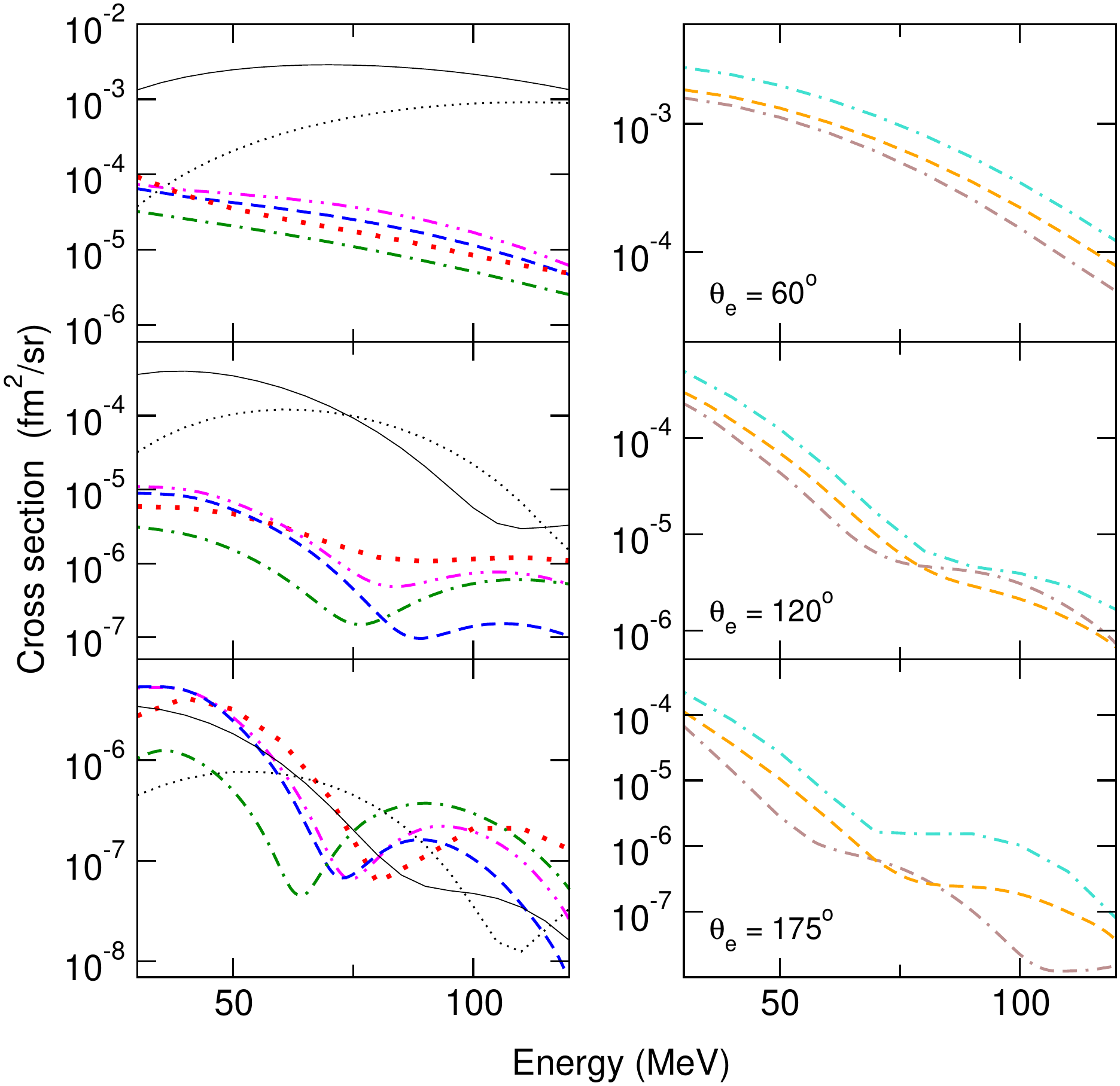}  
\end{center}  
\caption{\label{Fig8}  
Differential cross sections (d$\sigma$/d$\Omega$) for the excitation  
of the selected one-phonon $1^-$ states in $^{140}$Ce at  scattering  
angles $60^o$ (top row), $120^o$ (middle row), and $175^o$ (bottom row)  
as a function of incident energy $E_e$ of the electrons.  
The definition of the lines is the same as in Fig.~\ref{Fig7}.  
}  
\end{figure}

DWBA calculations have been performed for all one-phonon $1^-$ states of
$^{140}$Ce with an excitation energy below 20~MeV (42 states in total).   
We have considered incident energies from 30 to 120~MeV  
and scattering angles from 40$^o$ to 180$^o$.   
Special attention is paid to incident  energies of 40, 70, and  
110~MeV. They correspond roughly to the energies which may be achieved with the 
present set-up at  the S-DALINAC after the beam passes the linac once, twice, and four times.  
The main results of the  calculations are presented in  
Figs.~\ref{Fig5}-\ref{Fig8}.  
  
With Figs.~\ref{Fig5} and \ref{Fig6} we continue the discussion on   
the role of the longitudinal   
and transversal mechanisms of excitation which was started in  
connection with Figs.~\ref{Fig1} and \ref{Fig4}.   
One state at 6.51~MeV belonging to the PDR (left column) and one at
15.50~MeV belonging   
to the GDR (right column) are considered in each of the figures.   
The dependence of the differential cross sections on scattering angle  
is presented in Fig.~\ref{Fig5} for the incident energies of 40, 70, and   
110~MeV.  
The dependence on incident energy is shown in Fig.~\ref{Fig6} for the  
scattering angles $60^o$, $120^o$, and $175^o$.  
The DWBA cross sections are plotted by solid thick lines.      
By artificially setting $J_{1,L}(r)=0$ or $\rho_{1}(r)=0$ one obtains  
the excitation of the states by pure Coulomb (dashed line) or electric (dotted  
line) mechanisms, respectively.   
  
The analysis of the results in Figs.~\ref{Fig5} and \ref{Fig6} yields   
conclusions similar to the ones drawn from the PWBA predictions   
in the previous section: the electric term   
in the excitation of the GDR plays the most important role only for  
 very backward scattering, while for the PDR it may determine the cross  
section already at $90^o$.   
However, for some kinematics the interference between the  
Coulomb and electric parts may be extremely important (see, e.g., right-top  
panel of Fig.~\ref{Fig5}).  
In our examples this interference has often a destructive nature.   
  
The DWBA results  in Figs.~\ref{Fig5} and \ref{Fig6}   
are also compared to the PWBA predictions (thin solid lines).   
Basically, the DWBA leads to a smoothing of the sharp structures in PWBA due to 
the folding procedure.  
Although in some cases (e.g., right-center panel of Fig.~\ref{Fig5})   
the agreement between the two approximations is rather good,     
in other cases they may disagree by an order of magnitude or more with each
other.  
  
Cross sections for the states marked with an asterix in Fig.~\ref{Fig2} are  
plotted as a function of  scattering angle (Fig.~\ref{Fig7}) and  
as a function of  bombarding energy (Fig.~\ref{Fig8}). 
The selected states have the largest $B(E1)$ values in the PDR (left column) 
and GDR (right column) energy regions. The cross sections for other one-phonon 
$1^-$ states   
look rather similar except at the largest $q$-values in the studied  
kinematical range.
  
  
Low incident energies and small or modestly large scattering angles  
provide the biggest cross sections for the excitation of the PDR in   
inelastic electron scattering experiments. How large they are, can be seen  
from a comparison with the excitation cross sections for the $2^+_1$ and $3^-_1$  
states (with excitation energies 1.596 MeV and 2.464 MeV, respectively) 
presented in the left panels of Figs.~\ref{Fig7} and \ref{Fig8}   
by  solid and dotted lines.  
One notices that the one-phonon $1^-$ PDR states are  
about two orders of magnitude weaker excited in $(e,e')$ reactions than 
the $2^+_1$ state  
in a wide range of kinematics under consideration.    
An exception is the backmost scattering. The main reason is that  
for  the excitation of the $2^+_1$ and $3^-_1$ states the transversal  
part  plays a marginal role even at $\theta_e \to 180^o$. 
It is sizeable, however,  for the PDR states.
But when changing  the scattering angle from $40^o$ to $180^o$, the PDR  
excitation cross sections drop  by  one to three orders of  
magnitude, depending on  the incident electron energy.       
  
  
  
\section{Fine structure of the PDR}  
  
In order to discuss the absolute values of the  $(e,e')$ excitation cross 
sections   
for the $1^-$ states in the PDR energy region, which one  expects to measure  
in an experiment, it is necessary to account for the fact that   
the one-phonon $1^-$ states, discussed in the previous sections, are  
embedded in more complex two-, three-, etc. phonon states. The excitation  
of the latter from the ground state is very weak as compared to the  
excitation of the one-phonon states, but their density increases rapidly  
with excitation energy. The interaction between the one-phonon and more  
complex states leads to a fragmentation of the strength carried 
by the one-phonon excitations into components from many states 
with more complex wave functions. 
In other words, we are dealing with the decay of the doorway one-phonon states owing to the  
interaction with more complex background states in the spirit of
Ref.~\cite{BBB}.
  
In the QPM this decay is implemented by describing excited states   
with a wave function which contains one-phonon (first term), two-phonon (second term),  
and higher components     
\begin{eqnarray}  
|\Psi_{\lambda \mu}^{\nu} \rangle &=&   
\left\{ \sum_i R_i(\lambda \nu) Q^+_{\lambda \mu i}  
+  
\sum_{\lambda_1 i_1 \le \lambda_2 i_2 }  
P^{\lambda_2 i_2}_{\lambda_1 i_1}(\lambda\nu)  
\right .  
\nonumber  
\\  
&\times&    
\left .  
[Q^+_{\lambda_1 \mu_1 i_1} Q^+_{\lambda_2 \mu_2 i_2}]_{\lambda \mu}  
+ \cdots  
\right \} |\Psi_{g.s.}\rangle  
\label{eq7}  
\end{eqnarray}  
where $Q^+_{\lambda \mu i}$ is the creation operator of a  
phonon with  multipolarity $\lambda$ and its projection $\mu$, and where    
$i = 1, 2, 3 \ldots$ is the ordered number of the one-phonon states for a given  
$\lambda$. The phonon operators act on $|\Psi_{g.s.}\rangle$ which is the   
wave function of the ground state of even-even nuclei, identified 
with the phonon vacuum.   
Multiphonon configurations are built up of phonons of different  
multipolarities $(\lambda_1, \mu_1)$, $(\lambda_2, \mu_2)$, coupled to the 
same $(\lambda, \mu)$ as the one-phonon term 
$$[Q^+_{\lambda_1 \mu_1 i_1} Q^+_{\lambda_2 \mu_2 i_2}]_{\lambda \mu} = 
\sum_{\mu_1 \mu2} \langle \lambda_1 \mu_1 \lambda_2 \mu_2|
\lambda \mu \rangle Q^+_{\lambda_1 \mu_1 i_1} Q^+_{\lambda_2 \mu_2 i_2}.$$  
The eigenenergies of the states described by the wave functions (\ref{eq7}), 
as well as the coefficients $R_i(\lambda \nu)$ and   
$P^{\lambda_2 i_2}_{\lambda_1 i_1}(\lambda\nu)$,   
are obtained by the diagonalization of the model Hamiltonian on the set of 
these  
wave functions. Since the model Hamiltonian is already prediagonalized on  
the QRPA level, one-phonon configurations do not interact with each other, 
but they  
mix in the wave function (\ref{eq7}) due to their interaction with the same  
set of complex configurations.   

The transition densities of the states (\ref{eq7}) have the form of (\ref{eq7})  
where phonon operators are replaced by transition densities of   
one-, two-, etc. configurations.  
Neglecting the transition densities of the complex configurations, the cross  
section for excitation of the $\nu$-th ($\nu=1, 2, 3, \ldots$) state   
(\ref{eq7}) in $(e,e')$ reactions can be written as:  
\begin{equation}  
\left(\frac{d\sigma}{d\Omega}\right)_{\lambda \nu} \propto  
\sum_{m_s,\mu} \left|\sum_i R_i(\lambda \nu)  
A_i(\lambda \mu m_s)  
\right|^2  
\label{eq8}  
\end{equation}  
where $A_i$ are the transition amplitudes for the $i$-th one-phonon state.  
  
The first QPM calculation with the wave function (\ref{eq7}) for the PDR 
states was performed for $^{140}$Ce in late 90-ies (see Fig.~2 in
Ref.~\cite{He97}) and compared to the results of one of the first NRF experiment
in which the fine structure of the PDR was observed.
The model Hamiltonian was diagonalized in the basis of interactive one-, 
and a limited number of two-, and three-phonon configurations. 
The basis of complex configurations was extended later in \cite{wf}:    
two- and three-phonon configurations were built up from the phonons 
with  multipolarities from 1$^{\pm}$ to 9$^{\pm}$ and were cut above 8.5~MeV. 
All 42 one-phonon 1$^-$ configurations (discussed in the previous sections) 
were included in order to account for the GDR contribution at low excitation 
energies.  
The diagonalization yields 1157 $1^-$ states $\nu$  below 8.5~MeV.  
We will use this set of states in the discussion of the PDR below.
  
The fragmentation process of the $B(E1)$ strength of the doorway 
one-phonon $1^-$  
states in the PDR energy region is demonstrated in the left part of  
Fig.~\ref{Fig2}. 
To guide the eye, we also present in Fig.~\ref{Fig2}~(bottom) the strength 
function
\begin{equation}
S(B(E1), E_x) \propto \sum_{\nu}\frac{
B_{\nu}(E1)}{(E_x-E_{\nu})^2+(\Gamma/2)^2}
\label{sf}
\end{equation}
of the distribution where $E_{\nu}$ are the eigenenergies of the states
(\ref{eq7}) and $B_{\nu}(E1)$ are their reduced transition probabilities.
The strength functions here are calculated with an artificial width 
$\Gamma = 0.1$~MeV and presented in arbitrary units.

The strongest states described by Eq.~(\ref{eq7}) in Fig.~\ref{Fig2}~(bottom)   
have  $B(E1)$ values which are almost one order of magnitude smaller 
than the doorway ones in Fig.~\ref{Fig2}~(top). 
For the predictive power of the present set of the QPM wave functions 
we refer to Fig.~2 in \cite{Ce-last}. 
It combines information on excitation of
the individual PDR levels in $^{140}$Ce as observed in $(\gamma,\gamma')$, 
$(p,p')$ and $(\alpha, \alpha')$ reactions in comparison with the
calculation of the corresponding reaction cross sections performed with this 
set.
Although it is not possible to establish a one-to-one correspondence between
experiment and theory, a comparison of the calculations for single   
excitations in three different reactions with the experimental results on 
an absolute scale shows good agreement \cite{Ce-last}.
Also, calculations and experimental NRF data are found in good agreement
concerning 
the degree of fragmentation and on the integrated strength, if 
the sensitivity limit of the experiments is taken into account \cite{wf}.
All together, it leads us to expect that employing the same set of wave
functions in the calculation of the $(e,e')$  
cross sections will provide realistic values for the excitation of the  
strongest levels in the experiment. 

\begin{figure}  
\begin{center}  
\includegraphics[width=8.5cm]{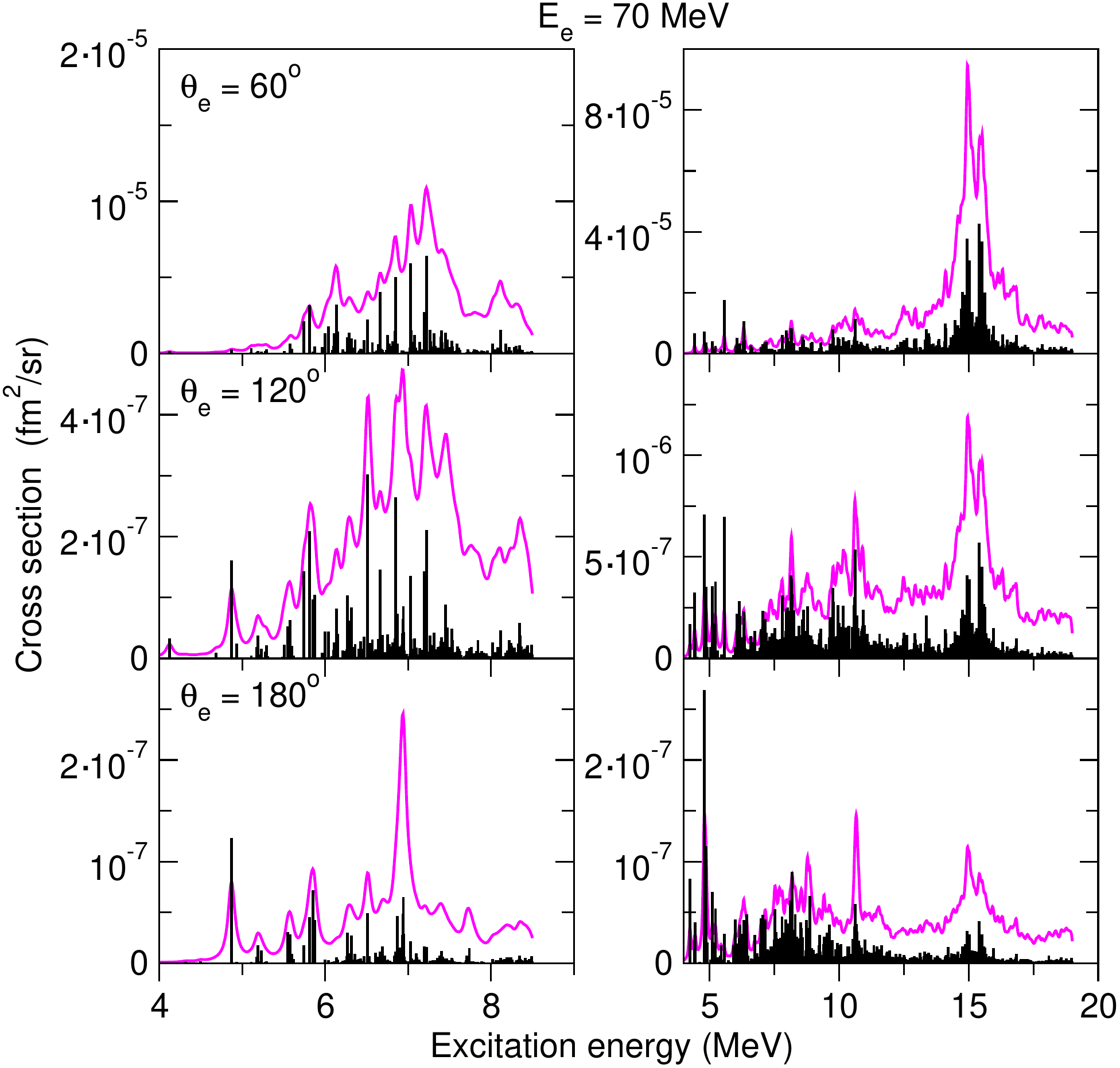}  
\end{center}  
\caption{\label{Fig10}  
Differential cross sections (d$\sigma$/d$\Omega$) for the excitation  
of the $1^-$ states in $^{140}$Ce   
by electrons with incident energy of 70~MeV  
at  scattering  
angles $60^o$, $120^o$, and $180^o$ (from top to bottom).   
The calculations are performed with the QPM wave functions (\ref{eq7})
which account for  
the coupling to complex configurations. Left column - the PDR, right column -  
the PDR and GDR.  
The strength functions are given by the smooth curves.
}  
\end{figure}

The results of the DWBA calculations with the QPM wave functions Eq.~(\ref{eq7})
are displayed in Fig.~\ref{Fig10} 
for an incident energy of 70~MeV and scattering angles $60^o$,  
$120^o$, and $180^o$. 
The strength functions in Fig.~\ref{Fig10} are defined similar to Eq.~(\ref{sf}) 
with the replacement of the $B_{\nu}(E1)$ quantities by the corresponding
$(e,e')$
cross sections $(d\sigma/d\Omega)_{\nu},$ and are presented in arbitrary
units which are different for different panels.
As in the case of the $B(E1)$ quantities, the largest cross sections 
of an individual $\nu$-th state in 
Fig.~\ref{Fig10} (left) are about one order  
of  magnitude smaller than the cross sections of the doorway  
one-phonon states 
for the same kinematics.   
  
To discuss the relative $(e,e')$ cross sections in the excitation of the PDR and GDR, 
an additional diagonalization of the QPM Hamiltonian has been performed 
by extending the basis of two-phonon configurations up to 19~MeV. 
No three-phonon configurations have been accounted for in this calculation.  
The fine structure of the GDR strength is shown in Fig.~\ref{Fig2} (right bottom) and  
Fig.~\ref{Fig10} for the $B(E1)$ values and $(e,e')$ cross sections,  
respectively.  
  
Figure \ref{Fig10} (left) 
demonstrates that, depending on the kinematics, the  
shape of the PDR excitation in inelastic electron scattering   
may vary dramatically and deviate from the distribution of the $B(E1)$ values  
which present the $q=0$ limit.  
For some kinematics the summed cross sections of all  PDR states 
from the $(e,e')$ reaction are even  
larger as compared to the summed ones for the GDR states. But the absolute 
values of the cross sections are small under such kinematical conditions.

\section{Conclusion}  
  
The excitation of the $1^-$ states in $^{140}$Ce by inelastically  scattered   
electrons with incident energies from 30 to 120 MeV is investigated.   
The scattering angle is varied from $40^o$ to $180^o$. This kinematical  
range covers a momentum transfer $q$ from 0.1 to 1.2~fm$^{-1}$.    
We consider  $1^-$ states which belong to the PDR and GDR. 
Their structure is described within the one-phonon QRPA, and by accounting 
for the coupling to  
complex configurations within the quasiparticle-phonon model.  
  
It is demonstrated that Coulomb   
scattering is the dominant excitation mechanism for the GDR states 
in an $(e,e')$ reaction in a  wide range of  scattering  
angles, except for the very backward scattering.  On the contrary, the PDR  
states are  predominantly excited by transverse electric scattering 
mediated by the nuclear current   
for scattering angles in a large angular region from $90^o$ to $180^o$.  
Also, the interference between the longitudinal and transversal components 
plays an important role for them. The latter effect is a distinctive feature of  
the DWBA calculations, while it is neglected in the PWBA.   
  
The calculations show that the fine structure 
of the PDR in $(e,e')$ reactions may change  
substantially, depending on the kinematics,  especially at large scattering  
angles. We predict that the $(e,e')$ excitation cross sections of the 
stron\-gest individual   
$1^-$ states are about three orders of magnitude lower than  
the respective cross section for the $2^+_1$ state,  except for very large  
scattering angles where the significant transversal contributions to the
cross section for the PDR dominates. However, the absolute values of the 
cross section are rather  
small.     
  
In this context we finally note that in earlier search for $M1$ and $M2$
giant resonances in $^{140}$Ce at the DALINAC the measured high-resolution
spectra -- $\Delta E$ varied between 28 and 48~keV (FWHM) -- at backward
angles showed no sign for excited $1^-$ states between excitation energies
from 7.5 to 10~MeV \cite{Meu}.
With the improved electron beams from the S-DALINAC and its high-resolution
spectrometers there is now, however, found hope to detect them. 

We thank Peter von Neumann-Cosel for discussions concerning the topic of
the present studies.
This work was supported by the  
Deutsche Forschungsgemeinschaft under Contract  
No. SFB-1245.  

\vspace*{2mm}
We have dedicated this article to our late colleague and friend Pier
Francesco Bortignon with whom we have discussed the physics of nuclear
excitations for many years.
  
%
%

\end{document}